# Some problems of Graph Based Codes for Belief Propagation decoding

## Quasi-cyclic Low Density Parity-check code (QC-LDPC)

A $(J,L)$ regular QC-LDPC code of length $N$ is defined by a parity-check matrix

$$H = \begin{bmatrix} I(p_{0,0}) & I(p_{0,1}) & \cdots & I(p_{0,L-1}) \\ I(p_{1,0}) & I(p_{1,1}) & & I(p_{1,L-1}) \\ \vdots & \vdots & \ddots & \vdots \\ I(p_{J-1,0}) & I(p_{J-1,1}) & \cdots & I(p_{J-1,L-1}) \end{bmatrix} \quad (1)$$

where $1 \leq j \leq J-1$, $1 \leq l \leq L-1$ and $I(p_{j,l})$ represents the $p \times p$ circulant permutation matrix obtained by cyclically right-shifting the $p \times p$ identity matrix $I(0)$ by $p_{j,l}$ positions, with $p = N/L$. For a specific QC-LDPC code we define the corresponding "base matrix" ("mother matrix") as the matrix of circulant shift that defines the QC-LDPC code:

$$B = \begin{bmatrix} p_{0,0} & p_{0,1} & \cdots & p_{0,L-1} \\ p_{1,0} & p_{1,1} & & p_{1,L-1} \\ \vdots & \vdots & \ddots & \vdots \\ p_{J-1,0} & p_{J-1,1} & \cdots & p_{J-1,L-1} \end{bmatrix} \quad (2).$$

Let define mask matrix for which regular QC-LDPC code can becomes irregular for different column weight case or QC-LDPC regular code with zero block circulant:

$$M = \begin{bmatrix} m_{0,0} & m_{0,1} & \cdots & m_{0,L-1} \\ m_{1,0} & m_{1,1} & & m_{1,L-1} \\ \vdots & \vdots & \ddots & \vdots \\ m_{J-1,0} & m_{J-1,1} & \cdots & m_{J-1,L-1} \end{bmatrix}. \quad (3)$$

$$\mathrm{H} = H \otimes M,$$

where $\otimes$ is Hadamar product

For $(J,L)$ QC-LDPC regular code mask matrix:

$$M_1 = \begin{bmatrix} 1 & 1 & \cdots & 1 \\ 1 & 1 & & 1 \\ \vdots & \vdots & \ddots & \vdots \\ 1 & 1 & \cdots & 1 \end{bmatrix}. \quad (4)$$

For $(J-1, L)$ QC-LDPC regular code with zero block matrix in first row mask matrix:

$$M_2 = \begin{bmatrix} 0 & 0 & \cdots & 0 \\ 1 & 1 & & 1 \\ \vdots & \vdots & \ddots & \vdots \\ 1 & 1 & \cdots & 1 \end{bmatrix}. \quad (5)$$

For $(J, L)$ QC-LDPC irregular code mask matrix:

$$M_3 = \begin{bmatrix} 0 & 1 & \cdots & 1 \\ 1 & 1 & & 1 \\ \vdots & \vdots & \ddots & \vdots \\ 1 & 1 & \cdots & 1 \end{bmatrix}. \quad (6)$$

In [1] proof that QC-LDPC code with parity-check matrix without zero block matrix $H = H \otimes M_1$ maximal value of girth equal 12.

Clear that mask matrix shall define column and row distribution of weight for QC-LDPC codes. Addition properties like easy encoding of repeat accumulate code can be easy add using special form of mask matrix, [2]. For example using:

$$M_{RA} = \begin{bmatrix} 1 & 1 & 1 & 1 & 1 & 1 & 0 & 0 & 0 & 0 \\ 1 & 1 & 1 & 1 & 1 & 1 & 1 & 0 & 0 & 0 \\ 1 & 1 & 1 & 1 & 1 & 0 & 1 & 1 & 0 & 0 \\ 1 & 1 & 1 & 1 & 1 & 0 & 0 & 1 & 1 & 0 \\ 1 & 1 & 1 & 1 & 0 & 0 & 0 & 0 & 1 & 1 \end{bmatrix} \quad (7)$$

we can generate irregular RA-codes with rate 0.5 and easy encoding properties.

Length of code in this case depends from size of circulant. This is why mask matrix is call protomatrix. Mask matrix approach often use for algebraic generated construction (EG/PG, BIBD, PBD, RS-based , [3-5] and another construction in Ramanujan graph class, [6-8]), to create irregular code from regular code, [3]. Main disadvantages of algebraic methods restriction on code parameters: length, rate, existence of linear dependent symbols and etc.

Mention that graph construction of code shall consist of two steps:

1. Choice degree distribution and define mask matrix;
2. labeling mask matrix – choice shift of integer permutation matrix for every '1' in mask matrix.

On the other hand graph can be constructed using random lifting methods (Guess-and-test, Hill-climbing, PEG) directly from protograph matrix(mask matrix, mother matrix),

[9-12]. These methods provide great freedom on code parameter choice but require greed search of best candidate in code ensemble (code lifted from similar protograph, mask matrix in term of algebraic construction) using multiple parameter optimization: ACE spectrum (TS enumerator), code-distance measure, Tanner Spectral properties, converge speed for protograph using EXIT-chart(mutual information), graph diameter and etc.

Reason of use Quasi-cyclic parity-check is efficient hardware implementation: high parallelism degree (inside CPM) which increases throughput and simplification wire-structures of decoder. For example QC-LDPC graph with authomorphism 42 (circulant size) enough to reach 6.7 Gbit/s with reasonable complexity (802.11ad) and more than 130 Gbit/s with full parallel implementation with magnitude grow of complexity, [13].

**Linear complexity BP decoder. Trapping set as a reason of BP fail**

If we consider some short length codes which we shall decode by maximum likelihood (ML) decoder we shall only try to improve codeword weigh spectrum.

For QC-LDPC codes of medium to long block lengths, the ML decoding is not suitable because of the enormous number of codeword's involved. Instead, some sub-optimal iterative decoding algorithms like the sum product algorithm (SPA) [14-15], min sum algorithm (MSA) [16] are devised. One significant assumption for which the SPA and MSA converges to the MAP decoding is that the states of all the check nodes (CN) connected to any specified variable node (VN) are independent given the value of the bit corresponding to the VN, [17]. Although this assumption holds asymptotically, it becomes invalid for finite-length codes after a certain number of iterations [17]. The reason for this is the presence of cycles in the Tanner graph. In a Tanner graph of girth $g$, the independence assumption becomes invalid after $m$ number of iterations where $m < \frac{g}{4} < m+1$, [17]. So, the girth of the Tanner graph should be increased in order to delay the violation of the independence assumption. This is reason of use low density parity-check codes with not impressive codeword's weight spectrum enumerator (even from first component- codewords with code distance weight) for BP decoding, Fig. 1.

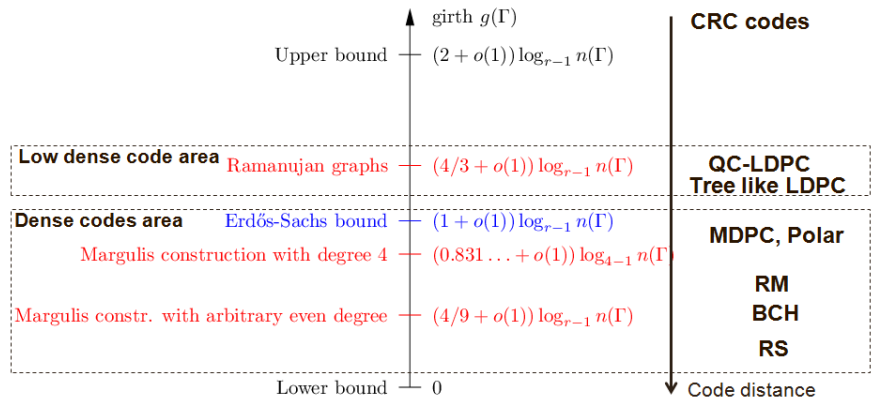

Figure 1. Asymptotical properties of graph girth and hamming distance grown from graph cardinally

Further research [18-19], show that cycles and their union form a special type of subgraph – Trapping set (TS), that become a reason of decoder failure under AWGN-channel.

Trapping set, TS$(a,b)$ is a sub-graph with $a$ variable nodes and $b$ odd degree checks. TS$(a,0)$ is just codeword of weight $a$. TS produced by union of cycles (which include all variable nodes in trapping set), Fig 2.

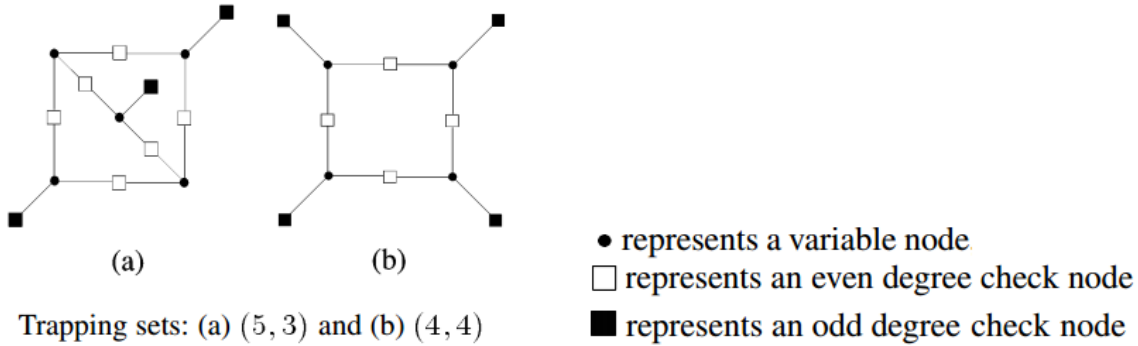

Trapping sets: (a) $(5,3)$ and (b) $(4,4)$

- represents a variable node
- ☐ represents an even degree check node
- ■ represents an odd degree check node

Figure 2. TS(5,3) and TS(4,4) representation

For examples TS(5,3) produced by three 8-cycles. TS(4,4) variable nodes involved in 8-cycles. Clearly than existence of g-size cycles (girth g) in Tanner graph fetch TS(g/2, g/2) or they overlap. Obviously that overlap of TS produce TS$(a,b)$, $b/a<1$ more harmful TS, because error on less number incident to odd degree parity-check variable nodes shall fetch failure on whole subgraph. This is why good code design methods must maximize value of TS$(a,b)$ threshold $b/a$. Indeed, if we have code with girth $g$, they don't have TS$(a,b): a+b<g$. It provide knowledge of TS structures for example girth 8 code, doesn't have TS(4,2), TS(3,3), very harmful TS(5,1) and low weight codeword TS(4,0) (it mean than minimal code distance is great than 4), [19]. But it is don't provide any detail about existence of another types of harmful TS ($b/a<1$, error in an $a$ VNs provide error in $b$ VNs) $a+b \geq g$.

Consider example: Margulis code with girth 8, [20]. From girth value we know that doesn't exist codeword with weight 4 TS(4,0), and all TS with $b<4$, but exist harmful TS(12,4) which for error in 4 VNs provide errors in 12 VNs, as result we shall have error floor at FER $10^{-6}$, Fig. 3.

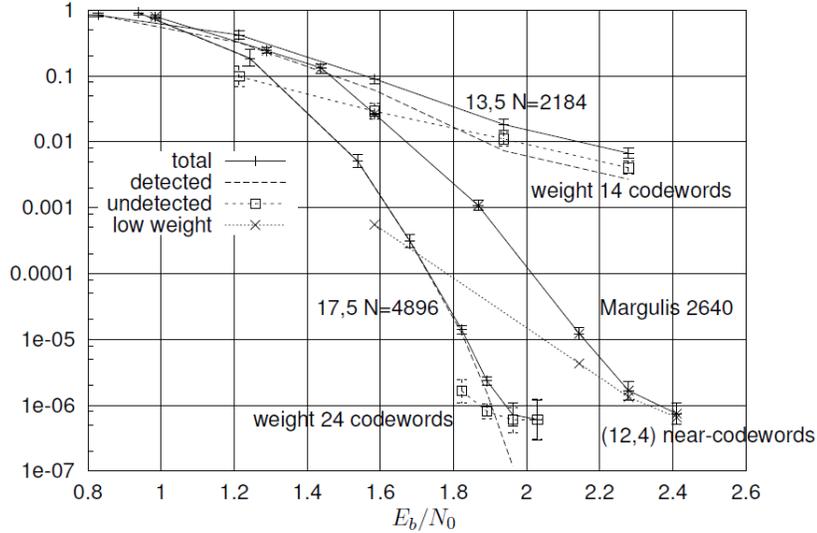

Figure 3. FER performance of codes on the AWGN channel under soft decoding 200 iterations.

**ACE Spectrum as measure of graph connective properties**

To measure quality of QC-LDPC code from TS characteristics let's define some special parameters Extrinsic Message Degree, it approximation version ACE and enumerator of ACE - ACE spectrum.

ACE spectrum is a spectrum of approximate cycle EMD metric. Extrinsic Message Degree (EMD) of the cycle, measures the level of connectively of the cycle with rest of the graph, how balanced TS structures, [21]. First of all, define several basic concepts about subgraph connective properties.

For a given cycle $C$ in the LDPC code graph let $V_c$ be the set of variable nodes in $C$ and $C(V_c)$ be the set of check node neighbours of $V_c$. We can divide the set $C(V_c)$ into three disjoint subsets:

- $C^{cyc}(V_c)$: subset of $C(V_c)$ belonging to the cycle $C$. Each node from $C^{cyc}(V_c)$ is at least doubly connected to the set $V_c$;
- $C^{cut}(V_c)$: subset of $C(V_c)$ that are not in the cycle $C$, but are at least doubly connected to the set $V_c$;
- $C^{ext}(V_c)$: subset of $C(V_c)$ singly connected to the set $V_c$.

For a given cycle $C$ in the code graph and the corresponding set $V_c$, let $E(V_c)$ be the set of edges incident to $V_c$. We can divide the set $E(V_c)$ into three disjoint subsets:

- $E^{cyc}(V_c)$: subset of cycle edges in $E(V_c)$ incident to check nodes in $C^{cyc}(V_c)$;
- $E^{cut}(V_c)$: subset of cut edges in $E(V_c)$ incident to check nodes in $C^{cut}(V_c)$;

- $E^{ext}(V_c)$: subset of extrinsic edges in $E(V_c)$ incident to check nodes in $C^{ext}(V_c)$.

Extrinsic Message Degree of a given cycle c in the code graph, denote $EMD(C)$, is $EMD(C) = |E^{ext}(V_c)|$, where $|E^{ext}(V_c)|$ is the cardinality of $E^{ext}(V_c)$, the number of extrinsic edges of $C$.

If a given cycle in code graph has low EMD, than its communication with the rest of the graph is limited. This limits the amount of new knowledge about values of VNs in the cycle that could be collected from rest of the graph. In the extreme case, when EMD of the cycle is zero, VNs in the cycles are isolated from rest of the graph and the cycle become low weight TS.

It is not an easy task to find EMD of the cycle in the graph, since it takes additional steps to determine if the edge is extrinsic edge or cut edge. If we neglect this difference and account for both, extrinsic and cut edges, into the cycle metric, we get simplified version of the EMD metric.

Approximated Cycle EMD(ACE) of a given cycle $C$ in the code graph, denoted ACE($C$) is $ACE(C) = |E^{ext}(V_c)| + |E^{cut}(V_c)|$. In this case $ACE(C)$ is easy to calculate:

$$ACE(C) = \sum_{v \in E(V_c)} (d(v) - 2), \; d(v) \text{ is the degree of variable node } v.$$

Let $G(H)$ be an code graph with $n$ variable nodes and let $d^{max}(v)$ be it largest left degree. Let $i$ be an even integers $4 \leq i \leq 2d_{max}$. For each $i$ let $\eta^i_{ACE} = (\eta^0_{ACE}(0),...,\eta^i_{ACE}(k_i - 1))$ be $k_i$-tuple of values where $\eta^i_{ACE}(i)$ is number of VNs with a property that smallest ACE Value of the cycle of length $i$ they belong to is equal to $i$, $0 \leq i \leq (k_i - 1) = (0.5)*(d^{max}_v - 2)$.

ACE spectrum of $G(H)$, $ACE(G(H)) = (n^4_{ACE},...,n^{2d\,max}_{ACE})$ is the $(d^{max}_v - 2)$-tuple of $n^i_{ACE}$ $k_i$-tuples, $4 \leq i \leq 2d_{max}$.

Consider simple example which show how increasing of ACE influence on TS, Fig. 4.

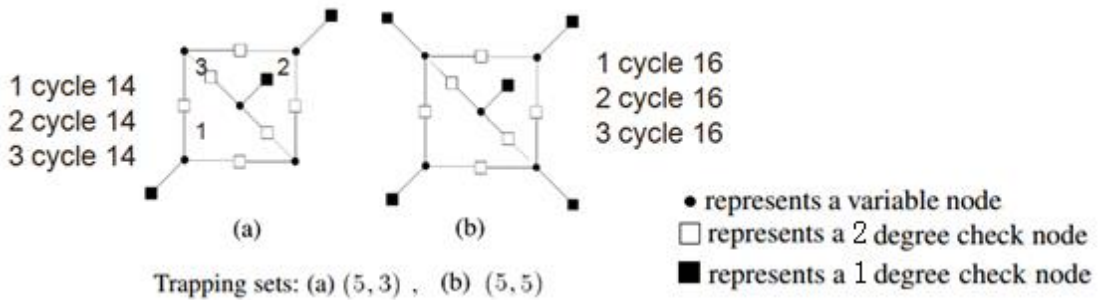

Figure 4. Calculated ACE value for Trapping set (5, 3) which consist of overlap between three cycles of length eight

Clear that increase of ACE value prevents harmful TS.

Mention that because approximate nature of ACE, if we use it as measure of TS we can leave some harmful TS. This why necessary to define sufficient parameters for which ACE spectrum measure, become EMD spectrum of QC-LDPC codes.

**Condition when ACE Spectrum becomes TS enumerator**

In paper [22] proof sufficient condition for equal EMD=ACE.

Theorem: For an LDPC code with girth $g$, any cycle $C_{2i}$ of length $2i$ has equal ACE and EMD if $i < g - 2$.

Example: For QC-LDPC code of girth 10, all the cycles of length $< 2*(10-2) = 16$: 10, 12 and 14 have equal respective ACE and EMD.

This leads us to two strong consequences, for any cycle with equal ACE and EMD:

1. CNs connected to the VNs in the cycle but not involved in the cycle will be singly connected to the cycle. The reliability of the messages coming from these CNs can be increased relative to messages coming from the CNs involved in the cycle in order to enhance the benefit of the messages coming from the rest of the Tanner graph. In this way, better connectivity can be ensured for the isolated small cycles and the performance of the iterative decoders can be improved (gain in waterfall region).

2. TS enumerators can be using easy obtain by ACE spectrum. Because for cycle of length $l$ contains $l/2$ number of VNs. For ACE value $\eta$, the subgraph induced by the $l/2$ number of VNs constrains exactly $\eta$ number of odd-degree CNs. Hence, the cycle can be treated as TS $(l/2, \eta)$, (error-floor region improve).

**Fast convergent properties**

From graph theory we have two important parameters which measure how fast messages dispersed to all nodes. This property is optional but can positively affect on performance under suboptimal BP-decoder.

As we already mention in chapter one we have $\lfloor (g-1)/4 \rfloor$ independent iterations, g – graph girth. Clear, that after $diameter/2$ iterations information from any VN shall have reached any other VN. It this case if $diameter < \lfloor g-1 \rfloor/2$ all vertex shall have statistically independent information. These considerations motivate to another heuristics of code on the graphs design with large girth and smallest diameter for the degrees of the vertexes, [23].

**Spectral properties for measure lower bound on pseudocodeword minimal weight**

Another property is described in [24]. Take bipartite graph defined by parity-check matrix with $r$ parity-check and $n$ bit-nodes then consider spectral graph (connective) property of graph by constructive real-value adjacency matrix $(r+n) \times (r+n)$:

$$A = \begin{pmatrix} 0 & H \\ H^T & 0 \end{pmatrix}$$

Diagonalize and take real eigenvalues: $\mu_1 > \mu_2 > ... > \mu_x$;

$$\omega_p^{\min}(H) \geq n \cdot \frac{2j - \mu_2}{\mu_1 - \mu_2},$$

$\omega_p^{\min}(H)$ – pseudocode word minimal weight [25]; $j, k$ - row and column weight of regular code.

Bound can be easy generalized for irregular code too [26].

Code with minimal ratio second eigenvalue to first eigenvalue $\min \frac{\mu_2}{\mu_1}$ is better because code distance agreed with minimal weight Trapping sets pseucodeword under belief propagation decoder.

**Density evolution method and Density evolution with punctured CN (ME-LDPC)**

In previous part we consider how to construct base on some base-matrix(protograph). For algebraic construction it usually simple regular code for which you use some mask matrix (protograph). For random method we begin directly from lifting of protograph. How to choice better photograph?

Performance of LPDC codes under BP decoding methods strong depend from degree distribution of columns $\lambda$ and rows $\rho$ in parity-check matrix, Fig. 5. On the figure you can see variance of the noise $\delta$ for AWGN channel for ensembles of regular codes with different column weight. Threshold show maximum level of channel noise which is likely to be (because estimated under infinity three graph –without cycle) corrected by a particular ensemble using the message-passing algorithm. Unfortunately lower column weight codes (capacity approaching) have better water-fall but together with poor error-floor and from another hand high column weigh code with extremely low error-floor have thresholds far from capacity.

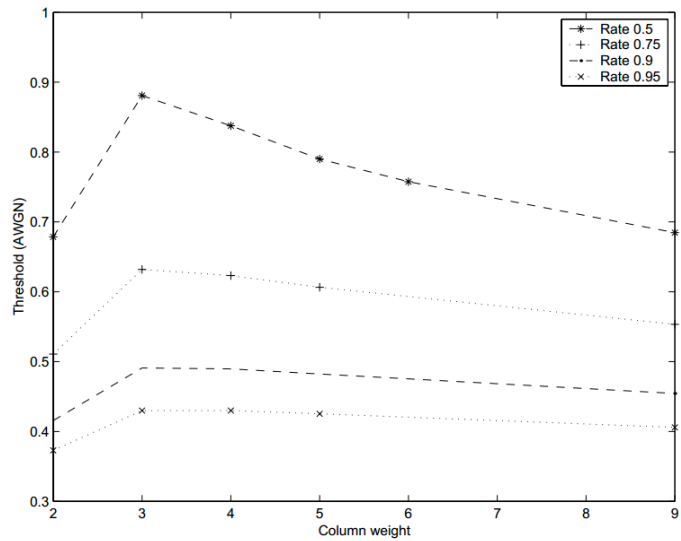

Figure 5. The threshold of regular LDPC codes on an AWGN channel calculated using density evolution, [27].

This is reason why low column weight VN must be compensating by high column weight VN. Requirement on the maximal column weight depend from value of gap from Shannon limit, Fig. 6, Fig 7. Another factor that effect on gap from capacity is quantization error, Fig. 7.

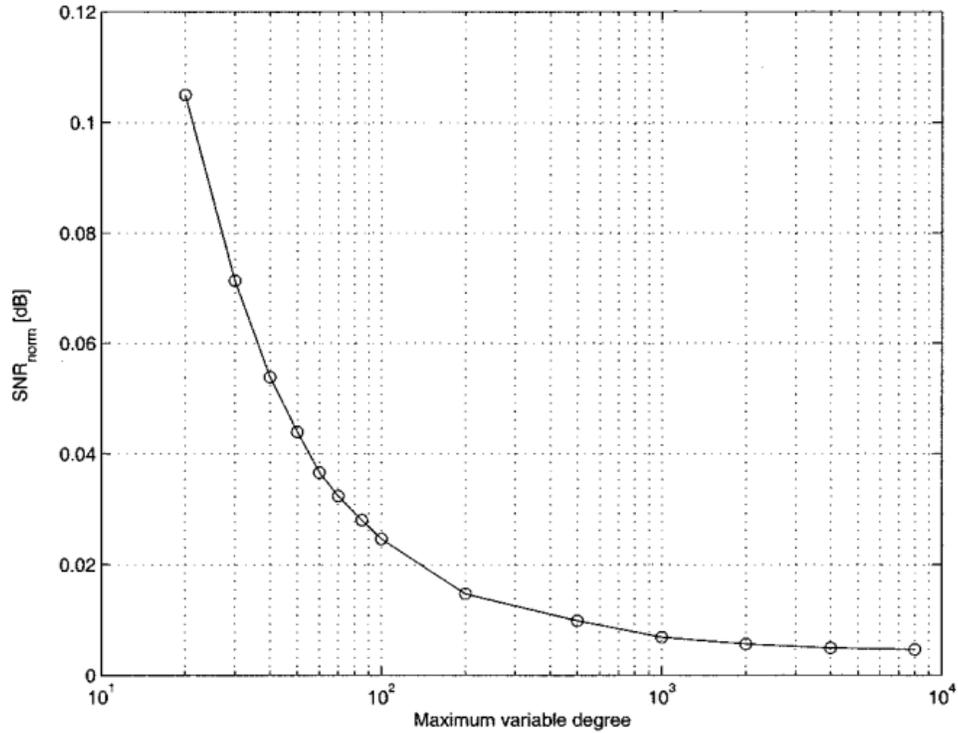

Figure 6. Threshold $SNR_{NORM}$ of rate ½ LDPC codes with maximum variable degree = 20…8000, [28]

| bits | threshold ($\sigma$) | error [dB] |
|------|----------------------|------------|
| 9    | 0.975122             | 0.01708    |
| 10   | 0.976918             | 0.00109    |
| 11   | 0.977025             | 0.00014    |
| 12   | 0.977037             | 0.00003    |
| 13   | 0.977040             | 0.00001    |
| 14   | 0.977041             | 0.00000    |

QUANTIZATION EFFECT

| $d_l$ | 100 | | 200 | | 8000 | |
|---|---|---|---|---|---|---|
| | $x$ | $\lambda_x$ | $x$ | $\lambda_x$ | $x$ | $\lambda_x$ |
| | 2 | 0.170031 | 2 | 0.153425 | 2 | 0.096294 |
| | 3 | 0.160460 | 3 | 0.147526 | 3 | 0.095393 |
| | 6 | 0.112837 | 6 | 0.041539 | 6 | 0.033599 |
| | 7 | 0.047489 | 7 | 0.147551 | 7 | 0.091918 |
| | 10 | 0.011481 | 18 | 0.047938 | 15 | 0.031642 |
| | 11 | 0.091537 | 19 | 0.119555 | 20 | 0.086563 |
| | 26 | 0.152978 | 55 | 0.036379 | 50 | 0.093896 |
| | 27 | 0.036131 | 56 | 0.126714 | 70 | 0.006035 |
| | 100 | 0.217056 | 200 | 0.179373 | 100 | 0.018375 |
| | | | | | 150 | 0.086919 |
| | | | | | 400 | 0.089018 |
| | | | | | 900 | 0.057176 |
| | | | | | 2000 | 0.085816 |
| | | | | | 3000 | 0.006163 |
| | | | | | 6000 | 0.003028 |
| | | | | | 8000 | 0.118165 |
| $\rho_{av}$ | 10.9375 | | 12.0000 | | 18.5000 | |
| $\sigma$ | 0.97592 | | 0.97704 | | 0.9781869 | |
| SNRnorm | 0.0247 | | 0.0147 | | 0.00450 | |

GOOD RATE-1/2 CODES WITH $d_l = 100, 200, 8000$

Figure 7. Influence of Quantization error and degree distribution for irregular LDPC codes, [28]

Density evolution use statistical considerations about fraction of edges that connect to variable and check nodes. More precisely we have two polynomials:

$$\lambda(x) = \sum_i \lambda_i x^{i-1}, \quad p(x) = \sum_i p_i x^{i-1},$$

Where $\lambda_i$ is a fraction of edges that connect to variables node of degree $i$ and

$p_i$ is a fraction of edges that connect to a check nodes of degree i.

$$\sum_i \lambda_i = 1, \sum_i p_i = 1.$$

Average variable and check degree can be expressed as:

$$l_{avg} = \frac{1}{\sum_i (\lambda_i / i)}, \quad r_{avg} = \frac{1}{\sum_i (p_i / i)}.$$

Code rate is defined $r = 1 - \frac{l_{avg}}{r_{avg}}$.

If $n \to \infty$, as I previously show graph become a computational tree. Then with probability $\lambda_i$ it has exactly $i$ neighbours and consequently it has one parent and i-1 children nodes. Each child of this variable node is a check node which in turn with probability $p_i$ has $i-1$ neighbours as it own children. In this case codes become completely defined from statically point of view. Consider a check node $c$ of this computational tree. Without constrains we will assume that its degree is at least 2, $i \geq 2$. Thus $c$ has a variable node $v$ as a parent node and it has $i-1$ variable nodes $v_s$, $s \in i,\dots,i-1$ as children nodes. Let $\gamma_s$ be log-likehood ration which is bit estimate known to a node $v_s$ in advance. In this case $\gamma_s$ are independent estimates. Run BP, the node $c$ received estimates $\gamma_s$ as message from its children nodes. Due to the fact that $c$ corresponds to parity-check equation to the later means that it obtains the estimations from all bits which participate in this equation except one which correspond to the parent node. $c$ process information from children nodes and estimates the value of the parent bit as follows:

$$m_{cv} = 2\tanh^{-1}\left[\prod_{s=1}^{i-1} \tanh(\gamma_s)\right],$$

And then send $m_{cv}$ as new estimates to the parent node $v$.

Independent random variables $\gamma_s$ with a density function $f$ have probability of $\gamma_s$ to be between $a$ and $b$ is given by

$$P(a < \gamma_s < b) = \int_a^b f(x)dx,$$

Then $m_{cv}$ is a random variable too. Density function $f_{cv}$ of $m_{cv}$ is equal:

$$f_{cv} = \sum_{i=2}^{i_{max}} p_i f^{(i-1)\otimes},$$

where $f^{(i-1)\otimes}$ is D-convolution.

Now let us consider a variable node of degree $i$ which belongs to the computational tree. It obtains messages from its children check nodes which are log-likehood estimates made by check node and make a decision. Denote this estimation by $\delta_s$. All messages are

statistically independent then a bit estimate $\delta$ made by this variable node is nothing but sum:

$$\delta = \sum_{s=1}^{i-1} \delta_s - \delta_0,$$

where $\delta_0$ is an initial channel estimate. Density evolution method implies that $\delta_s$ have the same density function $f$. Thus, for the $\delta$ density function $f_{cv}$ we can write

$$f_{cv} = f \times f \times .... \times f \times f_0 = f^{(i-1)\times} \times f_0,$$

×-convolution of density functions.

Computational tree variable nodes degrees are random integers associated with degree distribution $\lambda$.

$$f_{cv} = \left( \sum_{i=2}^{i_{max}} \lambda_i f^{(i-1)\times} \right) \times f_0.$$

The probability of erroneous estimate:

$$P_e = \int_{-\infty}^{0} f_{vc}(x) dx.$$

**Density Evolution:**

Let $f_0$ be a density function of initial log-likehood estimates; $I_\lambda$ be the maximum degree of variable nodes and $I_p$ be the maximum degree of check nodes.

1. $f_{cv} = f_0$
2. For $i$
3. $f_{cv} = \sum_{i=2}^{i_{max}} p_i f^{(i-1)\otimes}$
4. $f_{cv} = \left( \sum_{i=2}^{i_{max}} \lambda_i f^{(i-1)\times} \right) \times f_0.$
5. If $P_e = \int_{-\infty}^{0} f_{vc}(x) dx < \varepsilon$ then break
6. End for
1. If for any legibly value $\varepsilon$ there exist $i$ such that after $i$ iterations $P_e$ is less than $\varepsilon$ we will say that density evolution method converges.

Code with degree distribution $(\lambda, p)$ is optimal code if they maximal a decoding (BP) threshold $\sigma_+$:

$$(\lambda, p) = \arg \max_{\tilde{\lambda}, \tilde{p}} \sigma_+(\tilde{\lambda}, \tilde{p}).$$

We implemented method which is called differential evolution method, [29].

Differential evolution method. Start with a certain noise level $\sigma$. For the first generation $G = 0$ randomly choose NP degree distributions $(\lambda, p)$, $s = 1,..., NP$ (we will call it a list of distributions).

1. Initialization: for each distribution run $k$ step of the density evolution method and record its residual error $P_e$. Label the distribution with the smallest $P_e$ as the best distribution $(\lambda_{best}, p_{best})$

2. For the generation G+1 new distributions are generated according to the following scheme. For each $s = 1,..., NP$ randomly choose 4 distribution from the list with numbers be $s_1, s_2, s_3, s_4$, $1 \leq s_i \leq NP$ and $s_i \neq s$ and define new $(\lambda_s, p_s)$ as follows:

   2. $\lambda_s = \lambda_{best} + F(\lambda_{s_1} - \lambda_{s_2} + \lambda_{s_3} - \lambda_{s_4})$,
   3. $p_s = p_{best} + F(p_{s_1} - p_{s_2} + p_{s_3} - p_{s_4})$,
   4. where F is a real constant.

3. Selection scheme: for each distribution run $i$ step of the density evolution method and record its residual error $P_e$. Label the distribution with the smallest $P_e$ as the best distribution $(\lambda_{best}, p_{best})$.

4. Stopping criteria: if $P_{best}$ is not small enough go to the step 2. Otherwise, increase $\sigma$ slightly and return to step 1.

One is most disadvantages of methods that with large rate code $r = 1 - \frac{l_{avg}}{r_{avg}} > 0.7$ polynomial have large degrees and it is slow down speed of degree convergent. For more detail see, [30-31].

Could we do better protograph? Yes, Jeremy Thorpe from Robert McEliece team (McEliece try to create BP with correct marginal estimation, but get exponential grown of tree from iteration and begin consider how to improve threshold using puncturing VN in protograph, after show this methods for guy from JetPropulsion: Divsalar and etc). Generalized DE consider in [32]. One ultimate example of puncturing approach is factor graph design (based on cut-off boost phenomenon using Massey distraction) of Polar graph [33]. Just mention that pioneer research about improving of factor and bipartite graphs for soft decoder was previously considered in Forney's paper ([34]) and Wibergs's PhD thesis, [35].

**Problems:**

1. Can we construct protograph which combine: flexible lifting factor (easy change of length, start lifting value give enough parallelism degree to get throughput), with fast converge properties (diameter less as it possible, $diameter = \lfloor girth-1 \rfloor /2$) and code distance doesn't constrained by minimal Trapping sets pseucodeword weight?
   **Example:** Peterson graph with diameter 2, and girth 5; or some cages.
2. Can we measure more elements (not only use bound and complexity measure of code distance) in TS and codeword weight spectrum using spectral properties of graph or another "simple graph or algebraic properties" (which estimate with polynomial or sub exponential complexity)?
3. **Improve BP decoding divergent using sequential graph traversals, to get tree-like code.** Let we have some defined code on the graph based on the middle or high dense parity-check matrix. Using some sequentially scheduler (graph traversal) under BP decoding we can get graph without cycles. We get sparse extended parity-check matrix from dense parity-check matrix. For Example, BP decoding of dense Polar code, [33]. Could we do it in parallel (independent way, without graph traversal overlap)? What maximal parallel degree, with some defined parity-check dense? What tradeoff we can get between throughput and parallelism?
4. **Improve BP decoding divergent using puncture of variable node.** On Fig. 8-11 you can see protographs and base-matrix for Thorpe's protograph[32], AR4JA ([36]), $E^2RC$, [37] and Richarson's Multi-Edge LDPC [38].

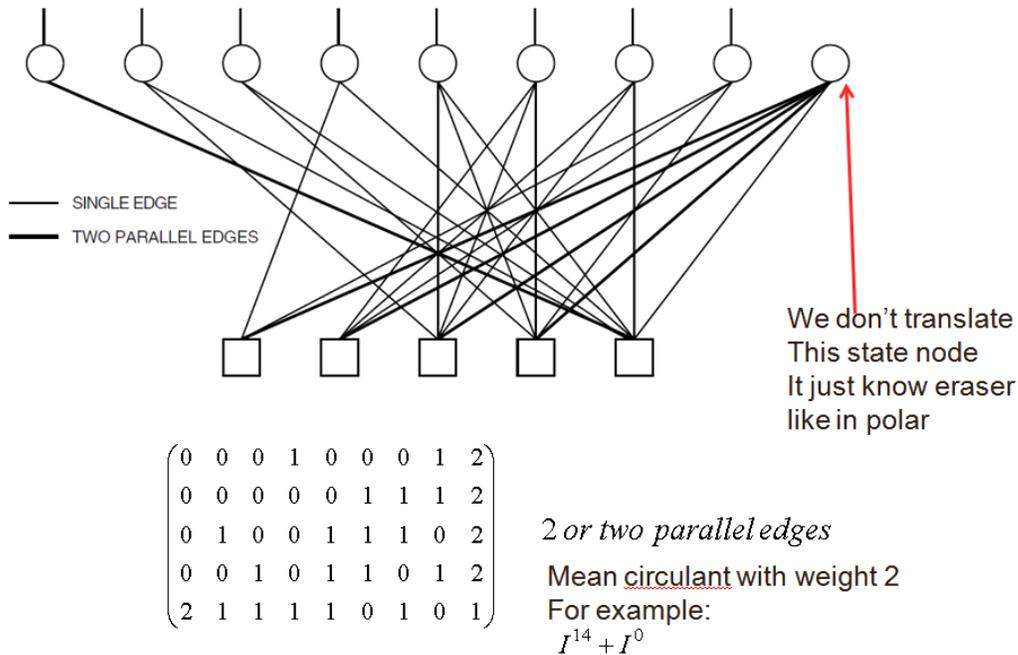

Figure 8. Thorpe's protograph based LDPC, last VN - state node

$$H_{4/5} = \begin{pmatrix} 1 & 0 & 0 & 0 & 0 & 0 & 0 & 0 & 0 & 0 & 2 \\ 0 & 1 & 1 & 1 & 3 & 1 & 3 & 1 & 3 & 1 & 3 \\ 0 & 1 & 2 & 2 & 1 & 3 & 1 & 3 & 1 & 3 & 1 \end{pmatrix}$$

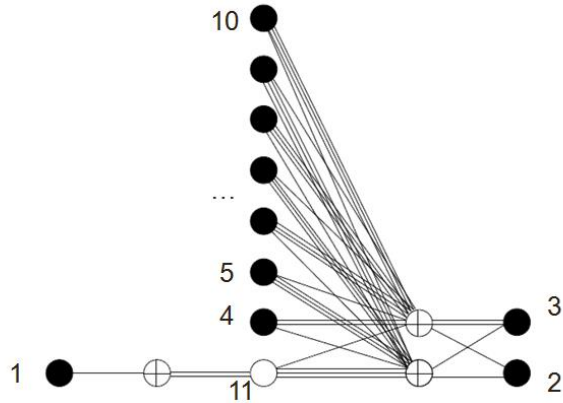

Figure 9. AR4JA protograph and base-matrix, 11 VN is puncture node (state node)

$$H = \begin{bmatrix} 2 & 0 & 1 & 1 & 0 & 1 & 0 & 1 & 1 & 1 & 1 & 0 & 1 & 0 & 0 & 0 \\ 4 & 2 & 0 & 1 & 0 & 1 & 0 & 1 & 1 & 0 & 0 & 0 & 1 & 0 & 0 & 0 \\ 3 & 0 & 0 & 0 & 0 & 0 & 2 & 0 & 0 & 0 & 1 & 0 & 0 & 1 & 0 & 0 \\ 3 & 2 & 1 & 0 & 2 & 0 & 0 & 0 & 0 & 0 & 0 & 0 & 0 & 1 & 0 & 0 \\ 3 & 0 & 0 & 0 & 0 & 0 & 0 & 0 & 1 & 0 & 1 & 0 & 0 & 0 & 0 & 1 \\ 3 & 0 & 0 & 1 & 0 & 1 & 0 & 1 & 0 & 0 & 0 & 0 & 0 & 0 & 0 & 1 \\ 3 & 0 & 0 & 0 & 1 & 0 & 1 & 0 & 0 & 0 & 0 & 1 & 0 & 0 & 1 & 0 \\ 3 & 4 & 1 & 0 & 0 & 0 & 0 & 0 & 1 & 0 & 0 & 0 & 0 & 0 & 1 & 0 \end{bmatrix}$$

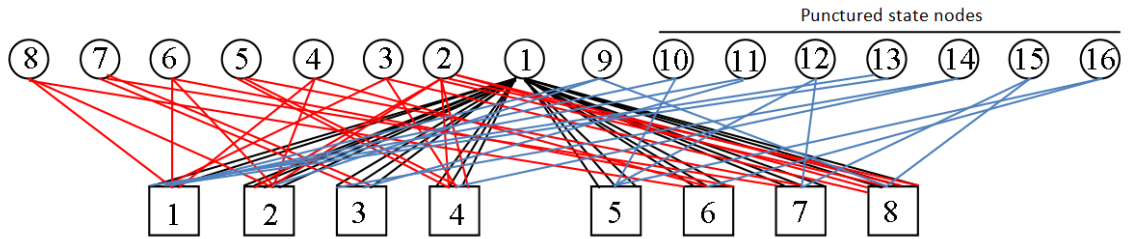

Figure 10. $E^2RC$ protograph(high rate optimized) and base-matrix, 10-16 VNs is puncture node (state nodes)

$$H = \begin{pmatrix} 1 & 0 & 0 & 0 & 0 & 0 & 0 & 0 & 0 & 0 & 0 & 0 & 0 & 0 & 1 & 0 & 1 \\ 0 & 1 & 0 & 0 & 0 & 0 & 0 & 0 & 0 & 0 & 0 & 0 & 0 & 0 & 0 & 1 & 0 \\ 0 & 0 & 1 & 0 & 0 & 0 & 1 & 1 & 0 & 1 & 0 & 1 & 0 & 1 & 0 & 1 & 0 \\ 0 & 0 & 1 & 1 & 0 & 0 & 0 & 0 & 1 & 1 & 1 & 0 & 1 & 0 & 0 & 0 & 0 \\ 0 & 0 & 0 & 1 & 1 & 0 & 0 & 1 & 1 & 0 & 0 & 1 & 1 & 1 & 1 & 0 & 1 \\ 0 & 0 & 0 & 0 & 1 & 1 & 0 & 1 & 0 & 1 & 1 & 1 & 0 & 1 & 1 & 1 & 0 \\ 0 & 0 & 0 & 0 & 0 & 1 & 1 & 0 & 1 & 0 & 1 & 0 & 1 & 0 & 0 & 0 & 1 \end{pmatrix}$$

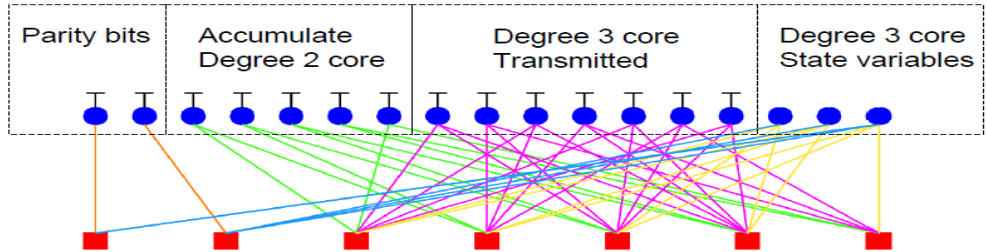

Figure 11. Richarson ME-LDPC

Remark(Standard approach to search protograph ensemble) :

We can get weight distribution from DE (with constrains from hardware implementation) use Theorem 8, [38] for search best code distance upper bound protograph ensemble(if circulant->infinity and we get convolutional code) and after use Theorem 7 to improve tightness, then using this upper bound estimate real code distance of our code. Then consider punctured nodes which improve divergent of graph under BP, [39]. But maybe we can found better general method?

Could we create new protograph with better divergent without greed search?

5. **Improve BP decoding divergent using addition parity-check from graph authomorphisms**. Let we have some defined code on the graph based on the middle or high dense parity-check matrix and its automorphisms representation. How many authomorphism representation of graph need to use under BP to probability (LLR) divergent under AWGN-channel, [41]? How to improve it using special protograph, with minimal matrix dense and special puncture structures?

6. **Improve BP decoding divergent using linearly dependent row in parity-check matrix**. Let we have some defined code on the graph based on the low, middle or high dense parity-check matrix and some linear depended row in parity-check. How to construct and minimize number of row to improve divergent under BP to probability (LLR), AWGN-channel, [42].

7. Could we lift protograph using different structure (not only by circulant permutation matrix(QC-LDPC)) and get with reasonable grow of complexity, better TS enumerator and properties of code (code distance, weight spectrum enumerator), without loss in throughput (save authomorphism of graph), [43-44] ?

Simple problem: If two step lifting, $1^{st}$ non-commuting, 2d QC, without greed search using permanent based estimation of code distance upper bound, [38-39]. More general problem if we use something different than QC or cycle code, could it be hardware friendly?